\documentclass{article}
\usepackage{floatrow}
\usepackage{spconf,amsmath,graphicx}
\usepackage{amssymb}
\usepackage{caption}
\usepackage{pifont}
\newcommand{\cmark}{\ding{51}}%
\newcommand{\xmark}{\ding{55}}%
\usepackage{arydshln}
\floatstyle{plaintop}
\restylefloat{table}

\title{Listen, Look and Deliberate: Visual context-aware speech recognition using pre-trained text-video representations}
\name{Shahram Ghorbani$^{12}$\sthanks{Work performed during an internship at Microsoft.},  Yashesh Gaur$^2$, Yu Shi$^2$, Jinyu Li$^2$}
\address{$^1$The University of Texas at Dallas, $^{2}$Microsoft Speech and Language Group}

\begin{document}
\ninept
\maketitle
\begin{abstract}
In this study, we try to address the problem of leveraging visual signals to improve Automatic Speech Recognition (ASR), also known as visual context-aware ASR (VC-ASR). We explore novel VC-ASR approaches to leverage video and text representations extracted by a self-supervised pre-trained text-video embedding model. Firstly, we propose a multi-stream attention architecture to leverage signals from both audio and video modalities. This architecture consists of separate encoders for the two modalities and a single decoder that attends over them. We show that this architecture is better than fusing modalities at the signal level.  Additionally, we also explore leveraging the visual information in a second pass model, which has also been referred to as a `deliberation model'. The deliberation model accepts audio representations and text hypotheses from the first pass ASR and combines them with a visual stream for an improved visual context-aware recognition. The proposed deliberation scheme can work on top of any well trained ASR and also enabled us to leverage the pre-trained text model to ground the hypotheses with the visual features. Our experiments on HOW2 dataset show that multi-stream and deliberation architectures are very effective at the VC-ASR task. We evaluate the proposed models for two scenarios; clean audio stream and distorted audio in which we mask out some specific words in the audio. The deliberation model outperforms the multi-stream model and achieves a relative WER improvement of 6\% and 8.7\% for the clean and masked data, respectively, compared to an audio-only model. The deliberation model also improves recovering the masked words by 59\% relative.

\end{abstract}
\section{Introduction}
\label{sec:intro}

Humans are inherently able to process and combine multi-modal information to understand language. For example, when we watch a narrated instructional video, we leverage all kinds of visual cues to resolves ambiguities we might have in understanding what has been said. Consequently, the counterpart problem of how to leverage visual information to improve ASR and natural language processing models has gained much attention in recent years. Previous works have demonstrated improvement by exploiting visual information for ASR \cite{2020looking,2018end,2017visual,2019analyzing,baseline,LM2018}, translation \cite{probing2019}, question-answering \cite{hori2019end}, dialog systems \cite{dialog} , and text summarization \cite{summarization}. 

It should be noted that Visual Context-aware Automatic Speech Recognition (VC-ASR) is different from the task of Audio-Visual Speech Recognition (AV-ASR) \cite{AV-ASR,lipreading}. AV-ASR can only be applied to videos in which the speaker's face information is available. However, it is not straightforward to apply such a model to open-domain videos. On the other hand, in VC-ASR, we leverage general-level visual information such as actions, objects, and places. This makes it suitable for open-domain videos but also much more challenging since the correlation between visual and speech signals in VC-ASR task is usually much smaller than AV-ASR.

For ASR, \cite{miao2016open} was one of the early works that investigated leveraging visual context to improve the modeling performance. They leveraged the objects and places features by early fusion, where they appended the visual features to the audio features, or normalized the audio features. This latter technique has also been referred to as visual adaptive training (VAT) in \cite{miao2016open}. Leveraging visual information for language modeling has been investigated in \cite{2017visual}, where they used the visual features as the beginning token of each sentence and demonstrated that the visual features have the potential to improve the ASR model at the language modeling level in addition to acoustic modeling. In \cite{baseline}, Ozan et al. attempted to exploit the visual features for a sequence-to-sequence (S2S) model using VAT or image captioning-inspired approaches that leverage the visual features to initialize the encoder and/or the decoder. However, in the aforementioned studies, the contribution of visual features is not clear. For example, Ozan et al. \cite{baseline} observed that the trained multi-modal model performs the same if the visual information is discarded during the inference. In our study, to validate whether the improvement is resulting from additional visual information or due to some regularization effects,  inspired by \cite{elliott2018adversarial}, we will also evaluate the VC-ASR models with random misaligned videos during inference time.

Nearly all the previous VC-ASR studies have extracted the visual features using models trained to classify pre-defined visual categories such as objects, places, and actions. However, the visual variability can not be covered by such a limited number of classes, and consequently, the features we get from these models are only representing the visual information partially. To address this problem, multi-modal joint pre-training \cite{miech2020end, li2020hero,JT,JT1} is shown to be a potential solution.  Miech et al. \cite{miech2020end} have pre-trained a text-video model to map the video and text to an embedding space where the similarity between text and video is high when they are semantically similar. In this study, we leverage this pre-trained model to extract rich visual information. Furthermore, as opposed to previous studies that average pool all the features of a video into a fixed vector, we extract a sequence of visual features as an input stream for the VC-ASR models.

Given two sequences of the audio and visual features, we propose to use a multi-stream attention-based S2S model to leverage these two inputs. In this model, which is an extension of an encoder-decoder based S2S  model \cite{s2s2,s2s1},  two attention heads are used to attend on audio and visual features to form a combined context vector.  In addition to this model, we also propose to leverage the visual context to post-process the N-best hypotheses generated by an audio-only model. For this purpose, we use a deliberation model architecture \cite{xia2017deliberation}. Previous studies have demonstrated the efficacy of the deliberation model for speech recognition \cite{hu2020deliberation,sainath2019two}, machine translation \cite{xia2017deliberation}, and speech-to-text translation \cite{sung2019towards}. For ASR, Hu et al. \cite{hu2020deliberation}, have shown improvement using the deliberation model on top of a streaming recurrent neural network transducer (RNN-T). In their study, the deliberation model had access to the future and past context and was able to refine the hypotheses produced by the first pass streaming model. In our study, we propose to leverage the visual context in a deliberation model trained on top of a pre-trained (audio-only) S2S model. The motivation behind this approach is that the deliberation can leverage the visual information to resolve the ambiguities in the N-best hypotheses and refine them to generate higher quality outputs. An important advantage of the deliberation model is that it has access to the N-best hypotheses generated by the first decoder.  Given the hypotheses, we can leverage the pre-trained text-video model to extract the visual context for each word in the hypotheses. Feeding the visual context along with the word embeddings to the deliberation model would help the model to resolve the ambiguities easier. This approach will be referred to as Visually Grounded Hypotheses (\textit{VG-Hypotheses}).

Since the speech modality is inherently a noisy medium, one practical scenario for which the visual information can be beneficial is where the speech modality is distorted. For VC-ASR, Tejas et al. \cite{2020looking} have investigated the utility of visual context where specific words are masked out in the audio modality. They concluded that the visual information is helpful to recover the masked words while the audio-only model is not able to. Inspired by this study, we also investigate how our proposed VC-ASR approaches perform when we replace some words in the audio modality with white noise.

In this study, we conduct the experiments on HOW2 data \cite{how2} which is collected from instructional videos on YouTube. We first examine the efficacy of the proposed multi-stream model to leverage the visual features for an improved VC-ASR model. The experiments demonstrate that this model performs 3.4\% relatively better than an audio-only model for clean data and 6.4\% for the masked data. The deliberation model trained on top of an audio-only model performs comparable with the multi-stream model, however, our analyses show that the deliberation model is superior in handling uncorrelated visual context. Furthermore, our results show that grounding hypotheses with the visual context (VG-Hypotheses) brings additional benefits in both the clean and masking scenarios. We hypothesized that VG-Hypotheses performs better when the N-best hypotheses from the first decoder are of greater quality. Therefore, we also train a deliberation 2nd pass model on top of a multi-stream audio-video 1st pass since it has better WER than the audio-only model. The resulting model achieves the best performance for the clean and masked data, with 6.0\% and 8.7\% relative WER improvement for clean and masked data, respectively, and 59\% relative improvement in recovering the masked words compared to an audio-only model.

\section{Visual features}
\label{sec:visualfeatures}
\begin{figure}[t]
  \centering
  \includegraphics[width=8cm]{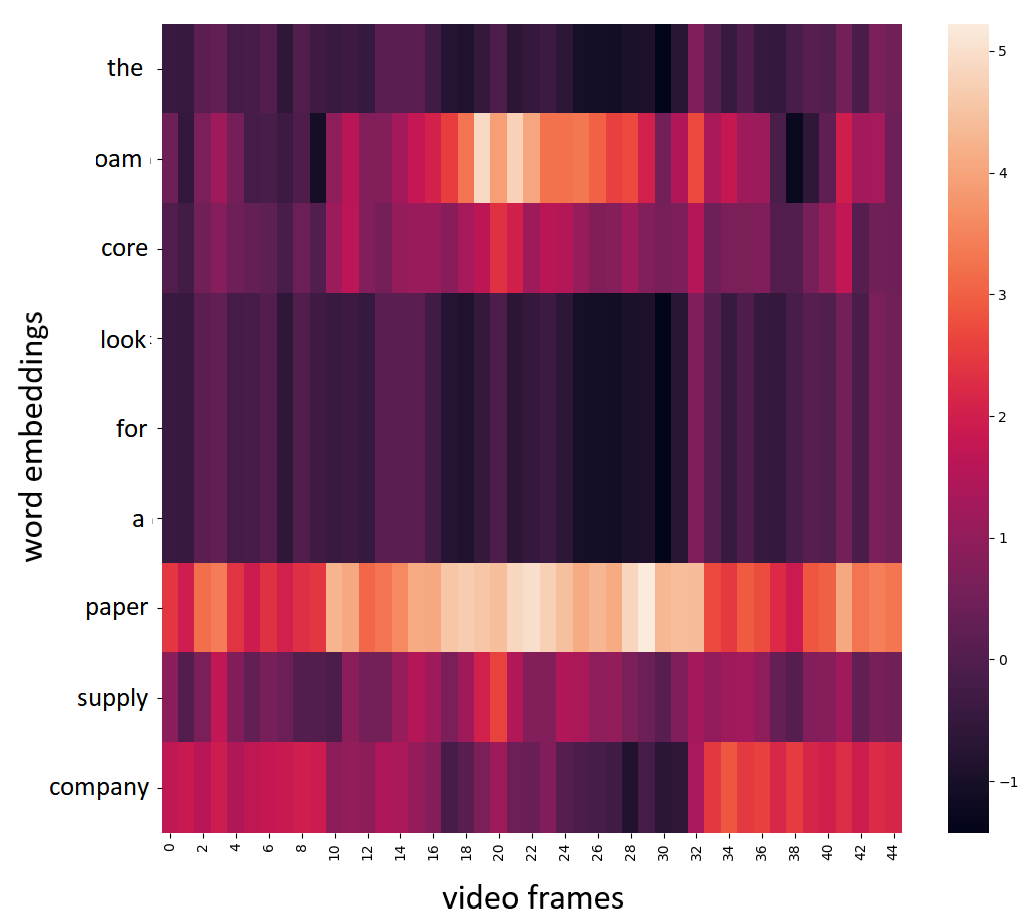}
  \vspace{-1em}
  \caption{A demonstration of the similarity matrix computed by multiplying word embedding (generated using the pre-trained text model) and visual features (generated using the pre-trained video model) of an utterance. }
  \label{fig:VF}
\end{figure}

To extract the visual features, we leverage a jointly pre-trained video-text embedding model \cite{miech2020end}. This model is trained to map 32 frames of video and its transcription to an embedding space where the similarity between these embeddings is high when they are similar semantically and is low otherwise. The detail of this model can be found in \cite{miech2020end}. Even though this model has been trained to map the entire sentence to an embedding, in our analysis we realized that the representation of each word has a high correlation with some parts of the video in which an object/action/place associated with that word appears, as illustrated in Fig.\ref{fig:VF}. This observation inspired us to use the pre-trained video model to map each video to a sequence of visual embeddings. The VC-ASR models in this study are designed to attend to this sequence of visual features to get the corresponding visual context for each word/token.

To match the training conditions of the pre-trained text-video model, for each video of $T \times H \times W \times 3$, first, we resize the frames resulting in a $T \times 200 \times 250 \times 3$ tensor. Then, the feature extraction is performed on a window of 32 frames with a sliding rate of 16 frames. The output of the pre-trained video model would be a $\frac{T}{16} \times 1024$ tensor.

\section{METHODOLOGY}
\label{sec:methods}

\begin{figure*}[htb]

\begin{minipage}[b]{.33\linewidth}
  \centering
  \centerline{\includegraphics[width=5.6cm]{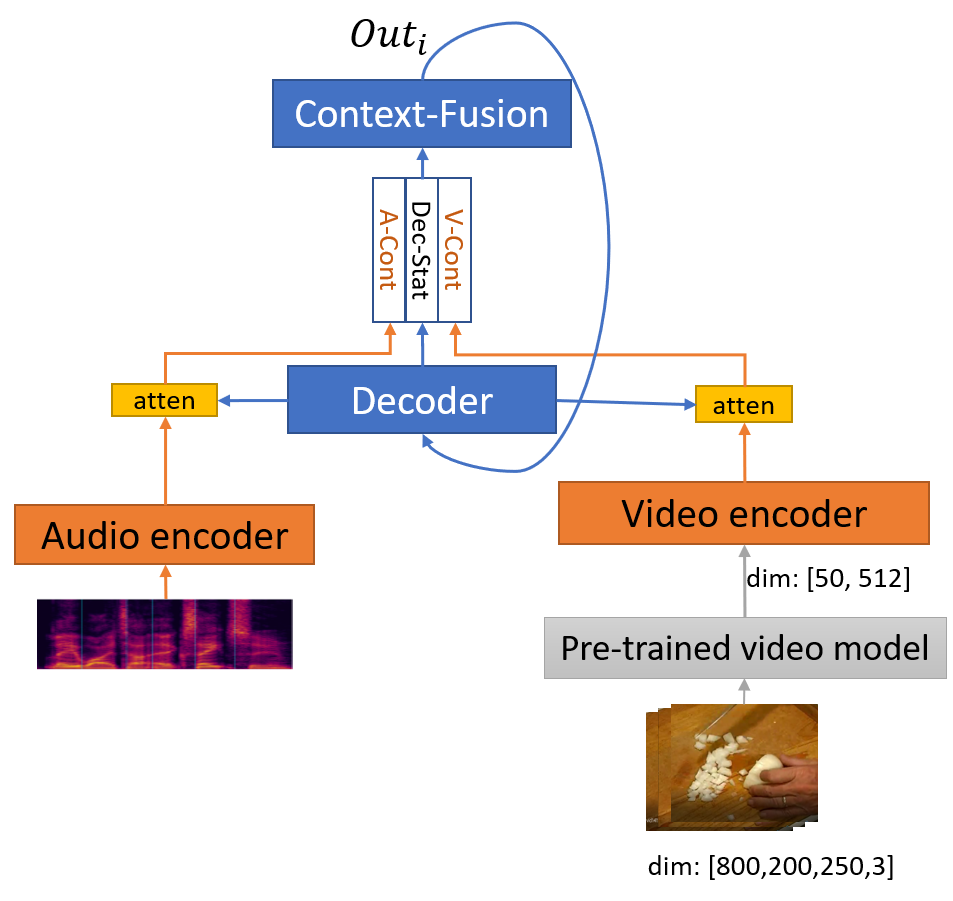}}
  \centerline{(a) Multi-stream model}\medskip
\end{minipage}
\begin{minipage}[b]{.33\linewidth}
  \centering
  \centerline{\includegraphics[width=5.6cm]{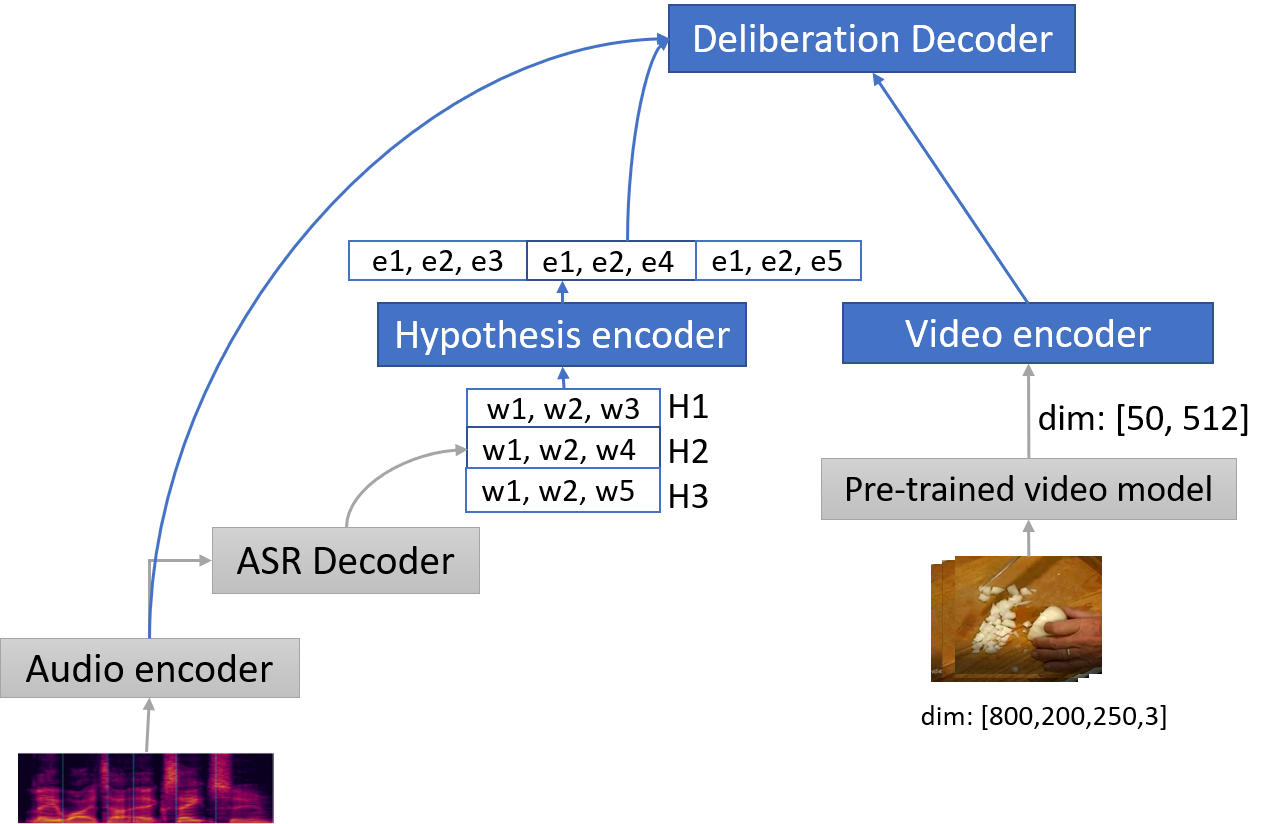}}
  \centerline{(b) Deliberation model}\medskip
\end{minipage}
\hfill
\begin{minipage}[b]{0.33\linewidth}
  \centering
  \centerline{\includegraphics[width=5.6cm]{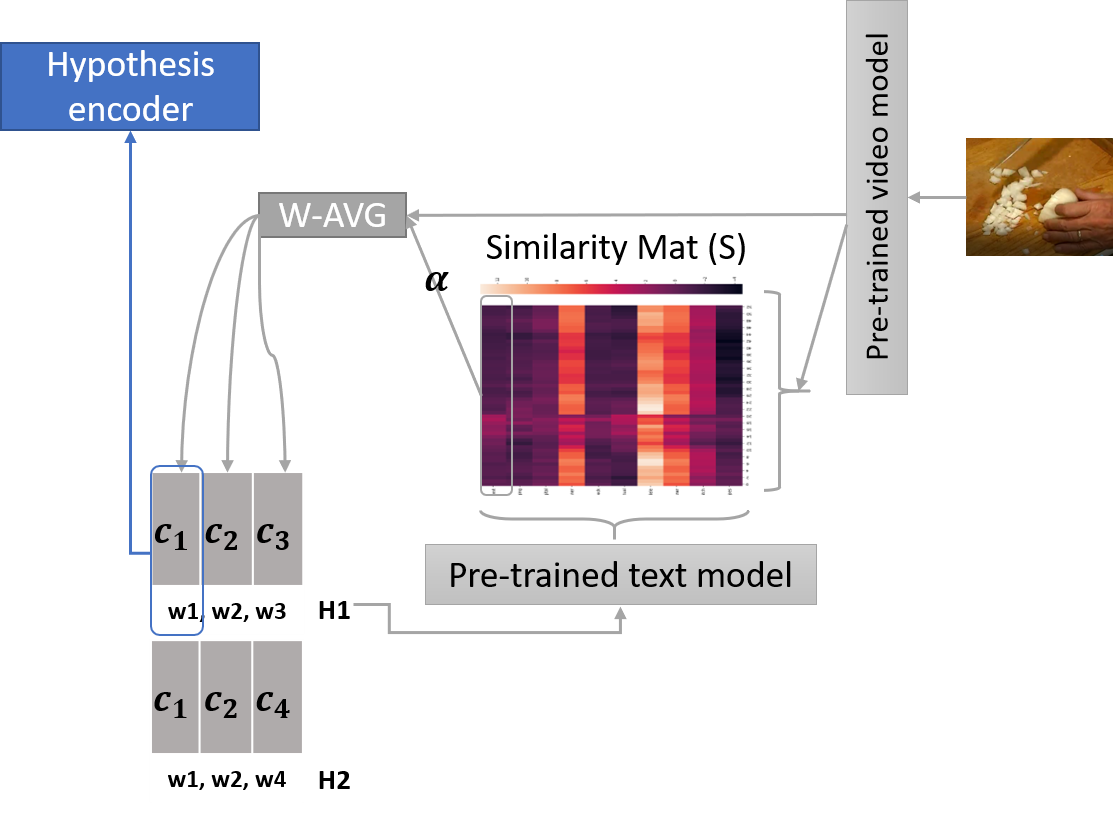}}
  \centerline{(c) Visually Grounded Hypotheses
  }\medskip
\end{minipage}
\caption{The proposed visual context-aware speech recognition models. For all the models, the visual features are extracted using the pre-trained video model.  The visual context features are then leveraged by one of three architecture. (a) \textbf{Multi-stream model}: This model extends an audio-only S2S model by adding a separate video encoder and an attention model to get the visual context. (b) \textbf{Deliberation model}: In this model, the visual features are leveraged in a  deliberation model on top of a trained (audio-only) S2S model. (c) \textbf{Visually Grounded Hypotheses}: This approach extends the deliberation model. For each hypothesis, the predicted words are used to attend on the visual stream to get the visual contexts. The resulting visual context vectors are then appended to the word embeddings before being passed to the hypothesis encoder. }
\label{fig:architectures}
\end{figure*}

In this section, we introduce our VC-ASR architectures. First, we describe our multi-stream S2S model and illustrate the detail of how this model processes and fuses the two input modalities. Then, we introduce the deliberation model and how we employ this model for the VC-ASR problem. These approaches are also illustrated in Fig. \ref{fig:architectures}.

\subsection{Multi-stream model}
\label{ssec:multi-stream}

In this approach, the model has two independent encoders to process the audio and video input streams. Then, two attention heads are used to attend on audio and visual encoded streams. At every step of outputting a token, the output of the previous decoder is used as a query to get the audio and video contexts.  Then these context vectors along with the decoder output are passed through a fusion network (denoted as Context-Fusion) that fuses these vectors into a combined context vector used to predict the next token. The S2S model outputs token probabilities which are matched to the ground truth tokens to train the model.  

We have investigated two architectures for the Context-Fusion;  concatenation and gating mechanism. 

\textbf{Concatenation (Cat)} -- In this architecture, given $\textbf{V}^C,\; \textbf{A}^C,\; \\ and\; \textbf{Dec}^O$ as the visual context, audio context, and decoder output, respectively, the concatenation approach is performed as follows, 
\begin{equation} \label{eq:1}
    Out_i =  (\textbf{W}^{cat}_1 Cat(\textbf{V}^C_i,\textbf{A}^C_i) + \textbf{b}^{cat}_1)  + \textbf{Dec}^O_i,
\end{equation}
where $\textbf{W}^{cat}_1 \in \mathbb{R}^{2d \times d} $ and $\textbf{b}^{cat}_1 \in \mathbb{R}^{d \times d}$ are parameters of a linear layer; and  $Cat$ stands for concatenating the context vectors on their last dimension that results in a combined context vector of size $2d$.

\textbf{Gating mechanism (Gate)} --  As we see in Fig. \ref{fig:VF}, for many words (e.g., stop words) of the utterances, there is no corresponding information in the video frames. Given the weak and noisy correlation between the transcriptions and the visual modality, for many of the decoding steps, the visual context does not provide any discriminative information for the decoder. On the other hand, for scenarios where the audio modality is distorted, the audio context would not be reliable too. Therefore, we propose to apply a gating mechanism to the audio and visual context to enable the model to ignore the noisy information from these modalities:
\begin{multline} \label{eq:1}
    \textbf{G\_V}^C_i =  \textbf{V}^C_i \odot \sigma(\textbf{W}^{gate}_1 Cat(\textbf{V}^C_i,\textbf{A}^C_i, \textbf{Dec}^O_i) + \textbf{b}^{gate}_1),\\ 
    \textbf{G\_A}^C_i =  \textbf{A}^C_i \odot \sigma(\textbf{W}^{gate}_2 Cat(\textbf{V}^C_i,\textbf{A}^C_i, \textbf{Dec}^O_i) + \textbf{b}^{gate}_2),\\
    Out_i = \textbf{G\_V}^C_i + \textbf{G\_A}^C_i + \textbf{Dec}^O_i,
\end{multline}
where $\textbf{W}^{gate}_1,\textbf{W}^{gate}_2 \in \mathbb{R}^{3d \times d}$, $\textbf{b}^{gate}_1,\textbf{b}^{gate}_2 \in \mathbb{R}^{d \times d}$ are learnable parameters of linear layers, $\sigma$ is an element-wise sigmoid activation and $\odot$ is the element-wise multiplication. 

\subsection{ Deliberation model}
\label{ssec:deliberation}

As shown in Fig. \ref{fig:architectures}, the proposed visual context-aware deliberation model consists of two steps. In the first step, the audio encoder and the first decoder process the audio modality to predict the target tokens. We first train these two components by minimizing the cross-entropy loss between the outputs of the first decoder and the ground truth tokens. Then, we freeze these components for the next step when we train the deliberation components. 

The encoded audio and the N-best hypotheses from the first decoder are used as input for the deliberation model. In addition to these two input streams, we also leverage the visual context at this step. The video frames are mapped to visual features using the pre-trained video model, then encoded by a separate video encoder. To process the hypotheses, following \cite{hu2020deliberation}, for each utterance, given the N-best hypotheses as $H_1, H_2, ..., H_N$, each hypothesis is processed separately by the hypothesis encoder, then the encoded hypotheses are concatenated in time to form an input stream for the second decoder. Similar to the multi-stream model, three attention heads are used to attend on audio, visual, and hypotheses encoded streams. We again examine the two Context-Fusion architectures, however, in this case, we have three context vectors including visual context, audio context, and hypothesis context. The resulting context vector from the Context-Fusion is used to predict the target tokens. In the second step of the training, we train the deliberation components to minimize the cross-entropy between the second decoder's outputs and the ground truth tokens.

 The proposed deliberation model has two important advantages compared to the multi-stream model introduced in section \ref{ssec:multi-stream}. First, this model introduces a flexible way to leverage the visual information on top of an existing audio-only model that can be trained with a large amount of data.  The amount of audio-visual-text paired data might be much smaller than audio-text paired data. In such a scenario, given a smaller amount of multi-modal data, we can leverage a pre-trained audio-only ASR model to produce the N-best hypotheses and the encoded audio representations, as input to a deliberation model. Then, we train a visual context-aware deliberation model to learn to leverage the visual context to refine the N-best hypotheses and generate higher quality outputs. 
 
\textbf{Visually Grounded Hypotheses} -- The second advantage of the deliberation model is that we also have a text modality input in form of the N-best hypotheses from the first pass audio-only model. It gives us an opportunity to exploit the correlation between visual and text representations as shown in Fig \ref{fig:VF}. For each hypothesis, which is a sequence of words, $w_1, w_2,..., w_n$, we leverage the pre-trained text model to map each word to a vector. Then, for word $w_i$ the visual context $\textbf{c}_i$ is computed as a weighted sum of the visual features  $\textbf{v}_1, \textbf{v}_2,..., \textbf{v}_m$,
\begin{equation} \label{eq:VC1}
    \textbf{c}_i = \sum_{j=1}^{m} \alpha_{ij} \textbf{v}_j,
\end{equation}

where $\alpha_{ij}$ for each visual vector $v_j$ is computed by
\begin{equation} \label{eq:VC2}
     \alpha_{ij} = \frac{\textbf{S}_{ij}}{\sum_{j=1}^{m} \textbf{S}_{ij}},
\end{equation}

where 
\begin{equation} \label{eq:VC3}
     \textbf{S}_{ij} = embedding(w_i) \cdot \textbf{v}_j + 2,
\end{equation}

\noindent where $embedding$ stands for the pre-trained text model representation. To make  $\textbf{S}_{ij} > 0.0$ for all cases, we added 2 to the result of the dot product in equation \ref{eq:VC3}. It should be noted that the dot product can effectively compute the similarity between the word embeddings and the visual features because the pre-trained text and video models are jointly optimized to maximize the dot product between text and visual representations of semantically similar pairs \cite{miech2020end}.

The resulting visual contexts are appended to the word embeddings before being passed to the hypothesis encoder. Appending these visual contexts can help the deliberation model to resolve the ambiguities in the N-best hypothesis. The mechanism to augment the N-best hypotheses with visual features is illustrated in Fig \ref{fig:architectures}. 
 
We hypothesize that since audio is the dominant modality for the VC-ASR problem, and also since the visual information is noisy and has a weak correlation with the target tokens, the multi-stream model, introduced in section \ref{ssec:multi-stream}, might not be trained well to get the optimum visual context from the input visual stream. We expect that the proposed deliberation model that uses the pre-trained text model to directly get the visual context can leverage the visual information better and consequently perform better. 

\section{Experiments and results}
\label{sec:experimets}

\subsection{Experimental Setup}

\textbf{Dataset \& features} --  All experiments are conducted on the HOW2 dataset \cite{how2} which is a collection of instructional videos from YouTube. This dataset provides around 185k (298 hours)  audio-visual utterances for training, 2022 utterances (3 hours) for development, and 2305 (4 hours) for evaluation. The official dataset comes with the audio features and action-based visual features for the three splits.  To follow the same experimental setting used in the baseline paper \cite{baseline}, we also leverage the provided audio and visual features.  For audio features, 40-dimensional filterbanks are extracted from time windows of 25ms with a 10ms frame shift. The resulting features are concatenated with 3-dimensional pitch features to form the final audio features. For the action-level visual features, a model trained to classify 400 different actions is leveraged \cite{hara2018can}. The action-based visual features are extracted from every 16 frames, then averaged pool into a 2048-dimension per each video. It should be noted that each video is split into smaller segments to create the utterances. However, the same video-level feature vector represents the visual features for all utterances of a video. These visual features are used for the baseline experiments. However, for the proposed models, we use the pre-trained text-video model described in section \ref{sec:visualfeatures} to extract a sequence of visual features per each video. In our experiments, we observed that the VC-ASR models trained with video-level visual features were always superior to ones trained with utterance-level visual-feature. Therefore, for all experiments with the proposed  models, we only report results trained with the former. 

The ground truth transcripts are preprocessed by removing the punctuations and mapping numbers to their words. Then, a SentencePiece tokenizer \cite{sentencepiece} model with 5000 vocabulary size is trained to map each transcript to a sequence of tokens.

\textbf{Audio-only model} -- The audio-only baseline is an attention-based sequence-to-sequence model \cite{s2s2,s2s1}. The (audio) encoder consists of 6 encoding layers. Each encoding layer is composed of a bidirectional gated recurrent units (GRU) \cite{gru} with 512 hidden units for each direction followed by a layer normalization. The decoder consists of two uni-directional GRU layers with 512 cells in each layer. A location-aware attention mechanism is used between the encoder and the decoder.

\textbf{The VC-ASR baseline model} --  In previous studies, the best VC-ASR performance for HOW2 data was achieved by applying VAT with the action-level visual features \cite{baseline}. To apply VAT, we first pre-train the audio-only model to full convergence. Starting from this pre-trained model, we add a linear layer to project the visual features to the audio feature space, then we add the projected vectors to all audio features. The whole model, including the projection layer, is tuned until convergence. This model is referred to as \textit{feature-level} fusion in the result tables.

\textbf{Multi-stream model} -- This model extends the audio-only model by adding a video encoder that has 3 encoding layers. A single-head location-aware attention mechanism is also used for the video modality. For this model, we don't need to pre-train the audio-only part and the entire model is trained from scratch.

\textbf{Deliberation model} -- The model for the first step has the same architecture as the audio-only model.   The first decoder generates 10-best hypotheses which are padded with \textless EOS\textgreater \;  token to the same length. The hypotheses' tokens are mapped to a 512-dimensional vector using the frozen embedding layer of the first decoder, then encoded by the hypotheses encoder which has 2 encoding layers.  The visual features are also passed through a visual encoder with three encoding layers. The deliberation decoder consists of two uni-directional GRU layers with 512 cells in each layer. Attention on each input stream (i.e., audio, video, and first-pass hypotheses) is performed using separate and independent single-head location-aware attention mechanisms \cite{s2s1}.

All the models in this study are trained to minimize the cross-entropy loss between the model outputs and one-hot encoding of the ground-truth tokens. Label-smoothing regularization \cite{labelsmoothing} with the probability of 0.2 is also applied. To train the models we use Adam optimizer \cite{adam} with initial learning rate $4.0 \times 10^{-4}$. The learning rate gets halved after every 50k iterations. The final model is chosen by monitoring the model's performance on the development set.  In the inference time, the decoding is performed with beam width 10 without using any external language model. The performance of the models is examined by measuring the word error rate (WER).

\textbf{Masking Words} -- We augment the How2 dataset by masking out specific words in the audio input in order to compare the ability of these models to recover the masked words. For each utterance, we leverage the pre-trained video and text models to compute the similarity between words and the visual features (Fig. \ref{fig:VF} shows an example of this similarity matrix). Then, words with similarity $> 3.5$ are considered as candidates to be masked out from the audio stream. For the training data, for each utterance,  we randomly sample from the candidates but limiting the number of masked words to be less than 10\% of the total number of words in the utterance. For the development and test sets, since we want to fix the masked words, we sort the candidate words based on their similarity with the visual features and choose the top-10\% words. For each chosen word, we set the masked region in audio features with white noise. 

\subsection{Results}
In this section, we compare the performance of the proposed VC-ASR models with various audio-only and audio-visual baselines. The results for the clean audio and masked words settings are reported in Table \ref{tab:clean_basline}-\ref{tab:masked_deliberation}. The performance of the models with random misaligned videos is also reported in the tables. For the masking scenario, we also report recovery rate (RR) of the masked words which is computed as 
\begin{equation} \label{eq:1}
    RR =  \frac{\#correctly\; recovered\; masked\;words}{\#masked\; words\; in\; the\; test\; set\;} \times 100.
\end{equation}

\textbf{Baseline models} -- As shown in Table \ref{tab:clean_basline}, our audio-only model is performing better than both the audio-only baseline and the visual context-aware baseline that was reported in \cite{baseline}. Since we use the same data and feature extraction as the baseline, the improvement is mainly attributed to our S2S model's architecture. We consider our improved audio-only model as the new baseline for the VC-ASR experiments.  We do not achieve any improvement by applying feature fusion (VAT) with action-level features on top of our audio-only model. We hypothesize that applying VAT with action-level features is serving as a regularizer, as is also shown in \cite{baseline}, and since our audio-only model is already performing well, this type of regularization can not improve the model performance.

\begin{table}[!th]
\caption{\label{tab:clean_basline} {\it  WERs of the baselines and the multi-stream model for the clean audio scenario. The results are reported in 3 sections as follows. First, the results from the previous studies. Second, our audio-only and the early fusion approach using action-level features. Third, the multi-stream model with just the visual features extracted from the pre-trained text-video model.  }}
\begin{center}
\begin{tabular}{ p{0.4cm}   p{2.5cm}  p{1.2cm} p{1.0cm}  p{1.2cm}  p{1.0cm}}
    \hline
        ID& Model & Visual Features & Test WER(\%)&  Misalgnd Videos WER(\%)  \\
      \hline 
     &Audio-only \cite{baseline}& --&19.1 & --  \\
     \hdashline 
     &Feature-Fusion& Action-&&\\
      &(VAT) \cite{baseline}&based &18.0& --\\
	  \hline 
	 A1& Our Audio-only & -- &17.77& --  \\
	 \hdashline 
	 &Our Feature- &Action-&  & \\
	 &Fusion (VAT) &based&  17.95 & 17.93 \\
	 \hline 
	  AV1&Multi-Stream &Text- &  &   \\
	  & &Video & \textbf{17.16} & 18.64\\
	 \hline
\end{tabular}
\end{center}
\end{table}

\begin{table}[!th]
\caption{\label{tab:masked_MS} {\it  WERs and Recovery rates (RR) of audio-only and multi-stream models for the masked words scenario. }}
\begin{center}
\begin{tabular}{ p{0.7cm} p{2cm} p{1.0cm} p{0.7cm} p{0.7cm} p{0.8cm}}
    \hline
        ID &Model &Context Fusion  & Test WER  (\%) & \; \; \; RR(\%) & Misalgnd Videos WER(\%)  \\
      \hline
     A2\_M &Audio-only&  --&22.64 & 22.29& --  \\
     \hline
      AV2\_M&Multi-Stream&Gate&  \textbf{21.17} & 33.75& 29.3  \\
       &Multi-Stream& Cat &22.4 & 28.0 & 24.56  \\
    \hline
\end{tabular}
\end{center}
\end{table}

\begin{figure*}

\centering
\floatbox[{\capbeside\thisfloatsetup{capbesideposition={right,top},capbesidewidth=12cm}}]{figure}[\FBwidth]
{\caption*{\textbf{Ref}: you need a lot of \textcolor{blue}{paint} on your \textcolor{blue}{brush} \\ \textbf{Audio-only}: you need a lot of \textcolor{red}{pin} on your \textcolor{red}{project}\\ \textbf{VC-ASR}: you need a lot of \textcolor{green}{paint} on your \textcolor{green}{brush}
}\label{fig:test}}
{\includegraphics[width=3cm,height=2cm]{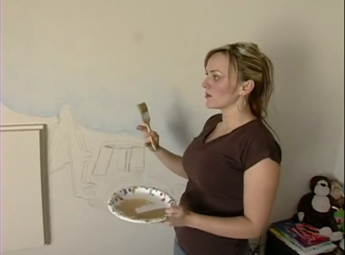}} 

\floatbox[{\capbeside\thisfloatsetup{capbesideposition={right,top},capbesidewidth=12cm}}]{figure}[\FBwidth]
{\caption*{\textbf{Ref}: there is not a whole lot to them they are not that robust they are mostly \textcolor{blue}{feathers} 
\\\textbf{Audio-only}: there's not a whole lot of you them they're not that room bust they are mostly \textcolor{red}{forward}\\
\textbf{VC-ASR}: there's not a whole lot you them they're not that robust they're mostly \textcolor{green}{feathers}
}\label{fig:test}}
{\includegraphics[width=3cm,height=2cm]{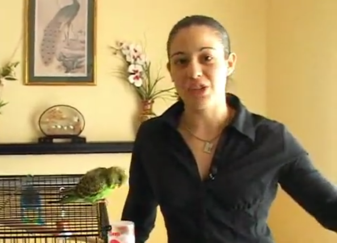}}

\floatbox[{\capbeside\thisfloatsetup{capbesideposition={right,top},capbesidewidth=12cm}}]{figure}[\FBwidth]
{\caption*{\textbf{Ref}: the key on this one and something to keep in mind on this \textcolor{blue}{position} is this\\
\textbf{Audio-only}: the key on this one and something to keep in mind on this \textcolor{red}{machine} is this\\
\textbf{VC-ASR}:\\Hyp1: the key on this one and something to keep in mind on this \textcolor{red}{jump} is this\\
 Hyp2: the key on this one and something to keep in mind on this \textcolor{green}{position} is this
}\label{fig:test}}
{\includegraphics[width=3cm,height=2cm]{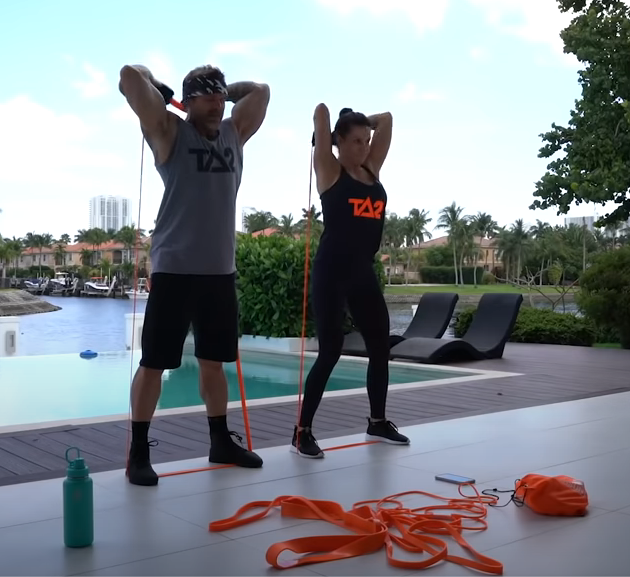}}

\end{figure*}

\begin{figure*}
\caption{Examples of how the audio-only and the multi-stream VC-ASR models perform for recovering the masked word. The blue words are the masked words that their corresponding region in the audio is set to white noise, the green words indicate the correctly recovered words, and the red words indicate the incorrect recovered words.}
\end{figure*}

\textbf{Multi-stream models} -- As is shown in Table \ref{tab:clean_basline}, for the clean audio data, leveraging the visual information in the multi-stream architecture improves the performance compared to the audio-only model (WER is improved from 17.77\% $\rightarrow$  17.16\%). To validate that the improvement is resulting from leveraging the visual information and not because of any regularization effects, the performance of the multi-stream model with misaligned videos is also reported. The results demonstrate that by feeding random misaligned videos the performance of the multi-stream model drops significantly and even becomes worse than the audio-only model. This observation confirms the effectiveness of the multi-stream model in leveraging the visual context.  It should be noted that an ideal VC-ASR model should leverage correlated visual context and disregard uncorrelated ones without hurting the performance. However, the multi-stream model did not show this behavior.

For the masked data, as is shown in Table \ref{tab:masked_MS}, the experiments demonstrate the superiority of the multi-stream model compared to the audio-only model. The multi-stream model achieves 6.4\% relative improvements in WER and 51.4\% relative improvement in the recovery rate. Here, the gating mechanism performs considerably better than the concatenation for fusing the context vectors from different modalities. We think this is because the gating mechanism can flexibly change the contribution of the audio and visual streams and compensate for the missing information in the audio stream better than the concatenation layer.  

\begin{table}[!th]
\caption{\label{tab:clean_deliberation} {\it WERs of the deliberation models for the clean audio scenario.}}
\begin{center}
\begin{tabular}{ p{0.5cm} p{0.5cm}p{0.8cm} p{1cm} p{1cm} p{1cm} p{1cm}}
    \hline
       \; \; ID &1st Pass&Visual Stream&VG-Hypthss&Context Fusion& Test WER(\%)&  Misalgnd Videos WER(\%)  \\
	 \hline
	
	 DB&A1&\; \xmark& \; \xmark& Cat & 17.58 & -- \\
	 \hline
	 D1&A1&\; \cmark& \; \xmark& Cat & 17.26 & \textbf{17.7} \\
	 
	  D2&A1& \; \cmark & \; \cmark&Cat & 17.07 & 17.87 \\
	 D3&A1&\; \cmark& \; \xmark& Gate& 17.33 &17.83 \\
	  D4&A1&\; \cmark &\; \cmark& Gate& 17.23 &17.91 \\
	 \hline
	  D5&AV1& \; \cmark&\; \xmark& Cat & 16.98 & 18.65\\
	  D6&AV1&\; \cmark&\; \cmark& Cat & \textbf{16.7} & 18.68 \\
	 \hline
\end{tabular}
\end{center}
\end{table}

\begin{table}[!th]
\caption{\label{tab:masked_deliberation} {\it   WERs and Recovery rates (RR) of the deliberation models for the masked words scenario.   }}
\begin{center}
\begin{tabular}{ p{0.8cm}p{0.9cm} p{0.8cm} p{1cm} p{0.7cm} p{0.8cm} p{0.9cm}}
    \hline
         1st Pass&Visual Stream& VG-Hypthss& Context Fusion  & Test WER  (\%)& \; \; RR(\%) & Misalgnd Videos WER(\%)  \\
    \hline
     A2\_M&\; \cmark&\; \xmark&Cat& 22.33      &25&    25.4\\
     A2\_M  &\; \cmark&\; \xmark& Gate& 21.06&  34&  23.19\\
    A2\_M&\; \cmark& \; \cmark& Gate& 20.94&33.5&\textbf{23.12}\\
    \hline
     AV2\_M &\; \cmark&\; \cmark& Gate& \textbf{20.65}&\textbf{35.5}& 24.11\\
    \hline
\end{tabular}
\end{center}
\end{table}

In Fig. 3 we consider a couple of examples of the test data where the multi-stream model performs better than the audio-only model. In the first two examples, the audio-only model is relying on the decoder's implicit language model to predict the masked word but it is unable to successfully recover them.  However, the multi-stream model correctly recovers the masked words by leveraging the visual context. In the third example, the multi-stream model is not able to recover the masked in the first hypothesis of the beam search, however, correctly recovers it in the second hypothesis.  An interesting observation in this example, which is also observed in many other utterances, is that the model substituted the target word ``\textit{position}" with the word ``\textit{jump}" which also has a high correlation with the visual context. It looks like the multi-stream model substitutes the masked words with words that are correlated with the visual context and have a high probability in the implicit language model. We validate this assumption by examining the quality of the N-best hypotheses which is measured by computing the best WER among the hypotheses (oracle WER). The oracle WER of the audio-only and multi-stream models are 18.9\% and 16.05\%, respectively. This means that the multi-stream model can recover many of the masked words in the N-best hypotheses. We will leverage this advantage in deliberation models.

\textbf{Deliberation models} -- As shown in Table \ref{tab:clean_deliberation}, for the clean data, the deliberation model outperforms the audio-only model and the multi-stream model (D2 vs A1 \& AV1).  However, to validate the contribution of the visual stream, we also examine a deliberation model without the visual stream (DB). The results confirm that the improvement is mainly attributed to leveraging the visual context rather than the deliberation process itself (D2 vs DB). For the masked data (Table \ref{tab:masked_deliberation}), the deliberation model performs slightly better than the multi-stream model (21.17 $\rightarrow$ 21.06). For Context-Fusion, the gating mechanism performs better for masked data and the concatenation performs slightly better for the clean data.

The results show that the deliberation model can disregard uncorrelated visual context better than the multi-stream model. For the misaligned videos, the deliberation model performs almost the same as the audio-only model. One possible explanation is that since the deliberation model processes the hypotheses generated by the audio-only model, it learns to attend more on the hypotheses as a reliable source, and leverage the visual context to resolve the ambiguities and correct the errors that exist in the hypotheses. Therefore, if the visual context does not bring any additional information this model would be able to perform close to the audio-only model.

\textbf{Visually Grounded Hypotheses} -- The experiments confirm that the VG-Hypotheses approach brings additional benefits in both the clean and masking scenarios. However, since VG-Hypotheses helps to resolve ambiguities among the candidates,  this approach performs better if the ground truth words are predicted in the N-best list.  Therefore, a model with lower oracle WER would be a better candidate for the first pass of the deliberation model.  The multi-stream models are performing significantly better than the audio-only models in terms of oracle WER. For the clean data,  the oracle WER for the audio-only model and the multi-stream models are 14.37\% and 12.48\%, respectively, and for the masked data, the results are 18.9\% and 16.05\%, respectively.  Therefore, we use the best multi-stream models, AV1 and AV2-M for clean and masked data, respectively, as the first pass.  The results of these deliberation models are reported in the last rows of Table \ref{tab:clean_deliberation} \& \ref{tab:masked_deliberation}.

The experiments support our hypothesis and we achieve the best performance for the clean and masked data using the deliberation with VG-Hypotheses on top of multi-stream models. For the clean data,  we achieve almost 6\% relative improvement, and for the masked data, we achieve 8.7\% relative improvement compared to the audio-only models.  The recovery rate for masked words is also improved by 59\% relatively compared to the audio-only model.

\section{Conclusion}
In this paper, we explored different approaches to leverage a pre-trained text-video representation for an improved visual context-aware ASR model.  We proposed multi-stream and deliberation models to leverage the visual features extracted by the pre-trained video model.  For the deliberation model, since we also have the text modality as input in form of  N-best hypotheses from the first pass, we leveraged the pre-trained text model to attend on the visual features directly. Experimental evaluation demonstrates the efficacy of the multi-stream and the deliberation models.  In addition to a clean audio stream scenario, we also examined how the VC-ASR models perform when the audio stream is distorted. For this purpose, we augment the data by masking out specific words in the audio stream.  The results show that the deliberation model performs better than the multi-stream model and achieves a relative WER improvement of 6\%  and 8.7\%  for the clean and masked data, respectively, compared to an audio-only model. The recovery rate of the masked words is also improved by 59\% relatively compared to the audio-only model.


\bibliographystyle{IEEEbib}
\bibliography{strings,refs}

\end{document}